\documentclass[11pt,twoside]{article}

\usepackage{asp2004}
\usepackage{epsf}
\usepackage{psfig}
\usepackage{lscape}
 
\begin{document}
\title{An Initial Look at the FIR-Radio Correlation within M51a using
  {\it Spitzer}}   
\author{Eric J. Murphy$^{1}$, Lee Armus$^{2}$, George Helou$^{2}$,
  Robert Braun$^{3}$, and the SINGS team}
\affil{$^{1}$Yale University; $^{2}${\it Spitzer} Science
  Center/Caltech; $^{3}$ASTRON, Netherlands}

\begin{abstract} 
We present an initial look at the FIR-radio correlation within the
star-forming disk of M51a using {\it Spitzer} MIPS imaging, observed
as part of the {\it Spitzer} Infrared Nearby Galaxies Survey (SINGS),
and WSRT radio continuum data.
At an estimated distance of 8.2 Mpc, we are able to probe the
variations in the 70$\micron$/22cm ratios across the disk of M51a at a
linearly projected scale of 0.75 kpc.
We measure a dispersion of 0.191 dex, comparable to the measured
dispersion of the global FIR-radio correlation. 
Such little scatter in the IR/radio ratios across the disk suggests
that we have yet to probe physical scales small enough to observe a
breakdown in the correlation.
We also find that the global 70$\micron$/22cm ratio of M51a is 25\%
larger than the median value of the disk, suggesting that the brighter
disk regions drive the globally measured ratio.
\end{abstract}

\section{FIR-Radio Correlation}
A major result of the IRAS all-sky survey was the discovery of an
empirical correlation between the far-IR (FIR) dust emission and the
optically thin radio continuum emission of normal late type galaxies
not containing an active galactic nuclei (AGN) \citep{de85,gxh85}.
Since, the correlation has been found to hold for galaxies spanning a
range of 5 orders of magnitude in luminosity, displaying a dispersion
of $\sim$0.2 dex. 
In an attempt to gain a better physical picture of the FIR-radio
correlation, we begin to look at its variations within the
star-forming disks of normal nearby galaxies being observed by {\it 
  Spitzer} as part of the {\it Spitzer} Infrared Nearby Galaxies
Survey (SINGS) \citep{rk03}.
Using circular apertures having diameters equal to the FWHM of the
70$\micron$ PSF, and assuming a distance to M51 of 8.2 Mpc, we probe
the disk at a linearly projected scale of 0.75 kpc.

\section{Results}
\begin{figure}[!ht]
  \centerline{
    \resizebox{8.7cm}{!}
	      {\plotone{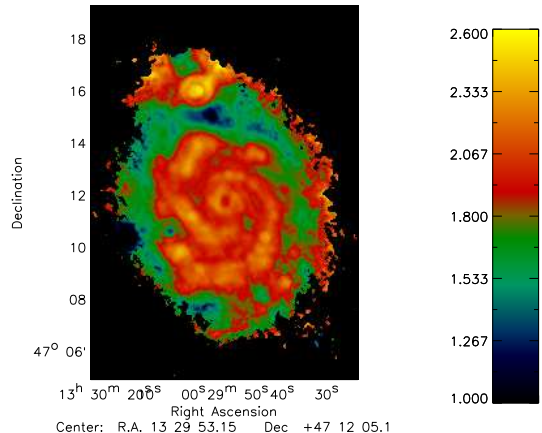}}
  }
  \caption{$q_{70} \equiv
    \log\left(\frac{f_{\nu}(70\mu{\rm m})}{f_{\nu}({\rm 22cm})}\right)$ map of
    M51 for pixels having 3$\sigma$ detections in both the input IR
    and radio images.\label{fig-1}}
\end{figure}

In figure \ref{fig-1} we plot $q_{70}$, the logarithmic ratio of
70$\micron$ infrared to 22cm radio emission, for M51.
The spiral structure of M51a is evident in the $q_{70}$ map showing
enhanced ratios along its arms with local peaks centered on HII
regions and depressed ratios located in the quiescent inter-arm and
outer-disk regions of the galaxy. 
Accordingly, we find a non-linearity of IR/radio ratios increasing
with IR surface brightness.
This non-linearity is consistent with what was reported by
\citet{mh95}, which is in fact opposite of the non-linearity observed
in the global FIR-radio correlation \citep{cox88}. 
We measure a global $q_{70}$ of 2.05 dex for M51a and find the median
and dispersion of $q_{70}$ for the disk to be 1.94 and 0.191 dex,
respectively.

\section{Conclusions}
From our initial look at the FIR-radio correlation within M51a, we
find that the measured dispersion is comparable to the dispersion in
the global FIR-radio correlation. 
This result was also observed by \citet{hoer98} who found the local
FIR-radio correlation within M31 to be as firm as the global
correlation on similar physical scales.
The small observed scatter indicates that we have not yet probed the
physical scales for which the correlation breaks down. 
This is consistent with the results of \citet{bp88} who found the
correlation to breakdown on the scale of a few hundred parsecs around
star-forming regions in Orion.
Last, we find the globally measured 70$\micron$/22cm flux ratio for
M51a is 25\% larger than the median value observed within its disk,
suggesting that the global ratio is weighted heavier by the brightest
regions.
A more detailed analysis of this work, including other galaxies,
can be found in our journal article \citep{ejm}. 



\end{document}